\documentclass[conference]{IEEEtran}
\usepackage{moreverb}
\usepackage{latexsym}
\usepackage{graphicx,subfigure}
\usepackage{times}
\usepackage{amssymb}
\usepackage{amsmath}
\usepackage{algorithmic}
\usepackage{algorithm}
\usepackage{amsthm}
\usepackage{comment}
\usepackage{textcomp}
\usepackage{xcolor}
\usepackage{multirow}

\usepackage{amssymb}

\usepackage{mathtools}

\usepackage[bookmarksopen,bookmarksnumbered]{hyperref}

\begin{document}
%
\title{\emph{IIDS}: Design of Intelligent Intrusion Detection System for Internet-of-Things Applications}
\author{
\IEEEauthorblockN{KG Raghavendra Narayan}
\IEEEauthorblockA{\textit{Computer Science and Engineering} \\
\textit{Indian Institute of Information Technology Sri City}\\
Sri City, Chittoor-517646, India.\\
Email: narayan.kgr@iiits.in\\
ORCID : 0000-0002-7905-863X}
\and

\IEEEauthorblockN{Srijanee Mookherji}
\IEEEauthorblockA{\textit{Computer Science and Engineering} \\
\textit{Indian Institute of Information Technology Sri City}\\
Sri City, Chittoor-517646, India \\
Email: srijanee.mookherji@iiits.in \\
ORCID : 0000-0001-8650-2463}
\and

\IEEEauthorblockN{Vanga Odelu}
\IEEEauthorblockA{\textit{Computer Science and Engineering} \\
\textit{Indian Institute of Information Technology Sri City}\\
Sri City, Chittoor-517646, India \\
Email: odelu.vanga@gmail.com \\ 
ORCID: 0000-0001-6903-0361}
\and

\IEEEauthorblockN{Rajendra Prasath}
\IEEEauthorblockA{\textit{Computer Science and Engineering} \\
\textit{Indian Institute of Information Technology Sri City}\\
Sri City, Chittoor-517646, India \\
Email: rajendra.prasath@iiits.in \\
ORCID: 0000-0002-0826-847X}

\and

\IEEEauthorblockN{Anish Chand Turlapaty}
\IEEEauthorblockA{\textit{Electronics and Communication Engineering} \\
\textit{Indian Institute of Information Technology Sri City}\\
Sri City, Chittoor-517646, India \\
Email: anish.turlapaty@iiits.in \\
ORCID: 0000-0003-0078-3845 }

\and

\IEEEauthorblockN{Ashok Kumar Das}
\IEEEauthorblockA{\textit{Center for Security, Theory and Algorithmic Research} \\
\textit{International Institute of Information Technology Hyderabad}\\
Hyderabad-500 032, India \\
Email: ashok.das@iiit.ac.in \\
ORCID: 0000-0002-5196-9589}
}

\maketitle

\begin{abstract}
With rapid technological growth, security attacks are drastically increasing. In many crucial Internet-of-Things (IoT) applications such as healthcare and defense, the early detection of security attacks plays a significant role in protecting huge resources. An intrusion detection system is used to address this problem. The signature-based approaches fail to detect zero-day attacks. So anomaly-based detection particularly AI tools, are becoming popular. In addition, the imbalanced dataset leads to biased results. In Machine Learning~(ML) models, $F_1$ score is an important metric to measure the accuracy of class-level correct predictions. The model may fail to detect the target samples if the $F_1$ is considerably low. It will lead to unrecoverable consequences in sensitive applications such as healthcare and defense. So, any improvement in the $F_1$ score has significant impact on the resource protection. In this paper, we present a framework for ML-based intrusion detection system for an imbalanced dataset. In this study, the most recent dataset, namely $CICIoT2023$ is considered. The random forest (RF) algorithm is used in the proposed framework. The proposed approach improves $3.72$\%, $3.75$\% and $4.69$\% in precision, recall and $F_1$ score, respectively, with the existing method. Additionally, for unsaturated classes (i.e., classes with $F_1$ score $< 0.99$), $F_1$ score improved significantly by $7.9$\%. As a result, the proposed approach is more suitable for IoT security applications for efficient detection of intrusion and is useful in further studies. 
\end{abstract}


\section{Introduction}
In recent years, Internet of Things (IoT) is widely using in many applications such as healthcare, defense, automation, and smart cities. By 2030, the expected increase in IoT devices worldwide would be approximately $29$ billions~\cite{report1}. In recent days, along with this IoT growth, the intrusion attacks are on the rise. For example, according to a recent report by NOKIA, around one million devices are involved in DDoS attacks in 2023~\cite{report2}. Therefore, design of a model for early detection of attacks becomes an emerging and challenging problem~\cite{verma2020machine}. In 1987,  Denning~\cite{denning1987intrusion} introduced the concept of Intrusion-Detection System~(IDS) that aims at early detection of possible attacks against information systems. After Denning's seminal work, there are many approaches, such as signature, statistical and anomaly-based, as presented in the literature~\cite{mukherjee1994, tavallaee2010, injadat2020, keserwani2021}. Comparatively, the anomaly-based approaches are more effective in detecting intrusion attacks, including zero-day attacks~\cite{liu2019,satam2020wids}. In anomaly-based detection, Machine Learning~(ML) and Deep Learning~(DL) models are attracting more attention. The efficiency of these models depends mainly on the class distribution in dataset~\cite{fernando2021dynamically, gumucsbacs2020comprehensive}. The IoT-based IDS datasets such as \emph{Bot-IoT}, \emph{IoT-23}, and $CICIoT2022$ contain various attacks that affect IoT networks~\cite{ciciot2023}. However, in most datasets, an extensive network topology with real IoT devices is not considered during data collection. The $CICIoT2023$~\cite{ciciot2023} is a real IoT attack dataset that includes an expansive topology which involves $105$ real IoT devices. The $33$ attack types are identified and grouped into seven high-level classes such as DoS, DDoS, Web-based, Recon, spoofing, Mirai, and brute force. Here, malicious IoT devices perform all attacks specifically targeting other IoT devices. In addition, the dataset encompasses various attacks not found in other IoT datasets. The benign data-capturing procedure focuses on collecting IoT traffic during idle states and with human interactions such as echo dot, sensor data, and smart camera video feeds. Since the dataset covered many possible attacks from the real-time network scenarios, studying the performance of various ML algorithms using such datasets facilitates better IDS solutions.

\par
In the IoT literature, there are many studies on various imbalanced datasets. In 2022, Elghalhoud $et$ al.\cite{elghalhoud2022data} used random oversampling (ROS) to balance datasets in their proposed model on \emph{BoT-IoT} and \emph{ToN-IoT}. Improvements are observed in various performance metrics such as recall, the $F_1$ score and Area Under Curve (AUC). Rashid $et$ al.~\cite{rashid2022anomaly}  presented another framework to study the imbalanced IoT datasets such as \emph{DS2OS} and \emph{Contiki} by using the Synthetic Minority Over-sampling. They observed an considerable enhancements in $F_1$ score. In 2023, Bowen $et$ al.~\cite{bowen2023blocnet} studied the imbalanced datasets including the IoT attack dataset \emph{IoT-23}. The notable improvements are observed in performance metrics $F_1$ score and recall. In conclusion, the application of class balancing techniques considerably improves the classification performance. 

\par 
In real-life applications, including IoT, security is a crucial and sensitive issue. Even a small improvement in early detection of any zero-day attacks  significantly impacts the resource protection. The performance of ML models is analyzed using the metrics such as Accuracy, Precision, Recall and $F_1$ score. Note that the $F_1$ score is an important metric in measuring the model accuracy in terms of class-level correct predictions. For example, we consider a data with $1000$ samples with $960$ non-target samples and $40$ target samples. Assume that a model predicts correctly the non-target samples. Suppose confusion matrix from the model is as follows: true negative (TN) is $960$, true positive (TP) is $1$, false negative (FN) is $39$ and false positive (FP) is $0$. Then, we can observe that the accuracy of the model is $96.1$\% and precision is $100$\% but recall is $2.5$\%, that is, the model can detect only the $2.5$\% of target samples. It is a very critical even though the model accuracy seems high but has extremely low recall and $F_1$ score ($4.9$\%). It indicates that the model is nearly failing to detect the target samples. Through the paper, we name the samples with less than $99$\% $F_1$ score as ``Unsaturated Classes (USC)''. In security applications, such as healthcare and defense, false detection of target samples may lead to unrecoverable consequences. Therefore, $F_1$ score is an important metric in the performance analysis of the model. Hence in this paper, we mainly consider the $F_1$ score to treat the imbalanced nature of the samples.

\section{Data Description and Observations} \label{s:dataset}
\begin{figure}[!htbp]
\begin{center}
\fbox{\includegraphics[width =\columnwidth]{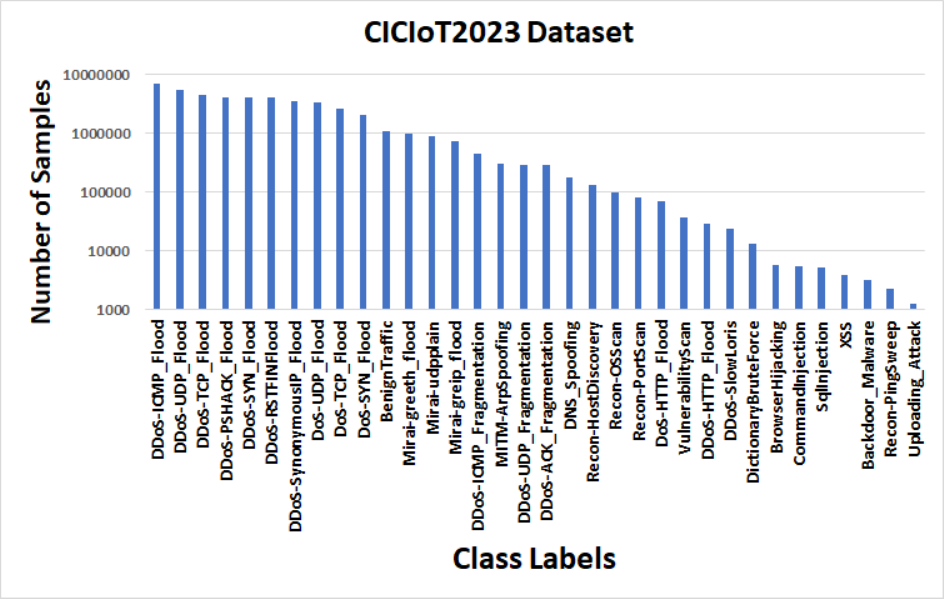}}
\end{center}
\caption{Class distribution of $CICIoT2023$ dataset} \label{fig:dataset-original-stat}
\end{figure}

The $CICIoT2023$ is a benchmark dataset designed by the Canadian Institute for Cybersecurity (CIC) to evaluate large-scale attacks within the IoT environment. The dataset is publicly available at the University of New Brunswick (UNB). It consists of $46686579$ samples collected from $105$ IoT devices. There are $33$ attack classes and one benign class in the dataset with $46$ features. The class-wise data distribution of $CICIoT2023$ is depicted in Fig.~\ref{fig:dataset-original-stat}. From the Fig. \ref{fig:dataset-original-stat}, it is observed that the dataset has skewed sample sizes. For example, the class $DDoS-ICMP\_Flood$ has the largest sample and the class $Uploading\_Attack$ has the smallest sample with a ratio of $5751:1$. When the training data has a skewed sample size distribution, it under-performs on the minority class instances~\cite{fernando2021dynamically}. So it is essential to perform data/class balancing to avoid biased results.

\begin{figure*}[!htb]
\begin{center}
\includegraphics[width =\textwidth]{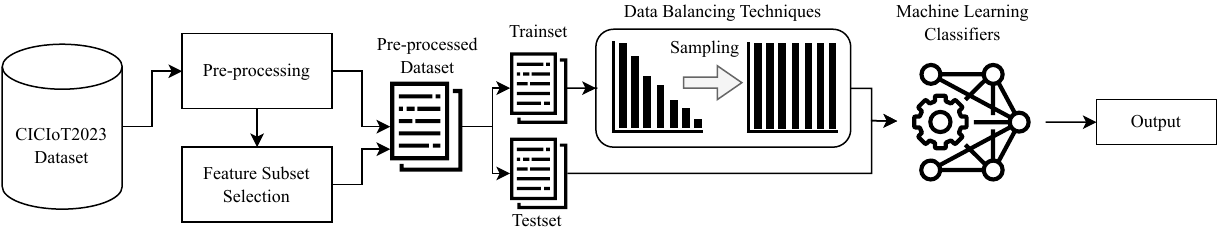}
\end{center}
\caption{Architecture of the proposed framework} \label{fig:architecture}
\end{figure*}

\section{Methodology} \label{sec:MainMethod}
{
The architecture of the proposed framework is depicted in Fig.~\ref{fig:architecture}. The proposed framework consists of three stages, namely (i). feature selection; (ii) class balancing; (iii) classification and assessment. The first stage consists of selection of an optimal feature subset from the dataset. In the second phase, the data is split into training and testing sets, and then data balancing techniques are applied on the training-set. Finally, in the third phase, the ML methods are applied on the pre-processed data and the classification performance is evaluated. The details of each stage of the methodology are discussed in the following. 
}

\textbf{Feature Selection}: In the first stage, feature subsets are selected using the following methods:
\begin{enumerate}
\item \textbf{CfsSubsetEval (CFS)}~\cite{hall1988correlation} using the Weka tool~\cite{hall2009weka}: It is a correlation-based feature selection method. Based on the predictability and the degree of redundancy of features the subset is assessed. Thus the features with higher correlation with the class labels and lower inter-correlation are selected. The first approach involves employing the CFS feature subset selection to obtain the top $k$ features from the existing feature set resulting in a subset of six features out of $46$ original features.

\item  \textbf{Intersection of RFE and MRMR methods (IRM):}  
    In this IRM approach, the $k$ best features through Random Forest (RF) with Recursive Feature Elimination (RFE) are selected. In the second step, another set of $k$ best features is chosen using the minimum Redundancy Maximum Relevance (mRMR)~\cite{peng2005feature} technique. Within these two subsets, the top $25$ features are chosen and an intersection of $11$ features is determined.
\end{enumerate}

\label{subsec:ClassBalance}
{
\textbf{Class Balancing}: In this stage, the selected dataset is pre-processed using sklearn (scikit-learn.org) standard scalar to normalise the dataset to unit variance. The $CICIoT2023$ dataset is split into train and test sets with $80:20$ ratio. In the analysis of the dataset, it is observed that there is a significant imbalance among the categories of attacks. Hence balancing techniques are deployed for improving the performance for minority classes without loosing performance for the major classes. 
The following class balancing techniques are applied on the train set. 
\begin{enumerate}
    \item Random Oversampling (ROS)~\cite{kubat1997addressing}
    artificially increases the number of samples in the minority classes. This process includes randomly duplicating instances from the minority class until it achieves the desired balancing with the majority classes. 
    \item Balanced Random Forest Classifier sampling (BRFC)~\cite{chen2004using}
    improves upon the standard Random Forest algorithm for handling imbalanced datasets. In a typical Random Forest algorithm, decision trees are trained on bootstrapped samples, which may favour the majority class in an imbalanced dataset. BRFC addresses this by adjusting the bootstrapping process to create balanced samples with an equal number of instances for both classes. This allows the ensemble to prioritize the minority class, enhancing performance on imbalanced datasets.
\end{enumerate}
}
{
The above sampling methods convert the train set of size $N_{tr} \times k$ into a balanced dataset $D$ of dimensions $N_R \times k$, whre $N_R$ is the sample size of balanced data. In Dataset $D$, each class has $N_L$ samples where $N_L = N_R/N_c$ where $N_c$ is the number of classes. 
}

{
\textbf{Classification and Assessment}: In the third stage, the balanced dataset $D$ is used to train the random forest models. The trained ML model is used to determine the class labels of the test set. A step-by-step process of the proposed framework is given in Algorithm~\ref{a:proposedalgo}. Different ML models and frameworks are evaluated based on the ML metrics given below. The metrics used in this research are : 1) Precision, 2) Recall, 3) $F_1$ score, 4) Accuracy and 5) Cohen Kappa score. 
}

\begin{algorithm}[!htbp]
\caption{The Proposed Intelligent Intrusion Detection System (IIDS)}{\bf Input:} \label{a:proposedalgo}
Dataset with $n$ features $D=\{ df_1, df_2, df_3, \cdots, df_n\}$ 

\smallskip
{\bf Description:}
\begin{algorithmic}[1] \small
 \FOR{$df_i$ $\in$ $D$}
    \STATE {Data Normalization: scaling to unit variance.} 
 \ENDFOR
 \FOR{$df_i$ $\in$ $D$} {
 \STATE Identify $k$ best features from $D$
 \STATE {Feed the data with selected features,\\ $SF=$ $\{ sf_1,$ $sf_2,$ $sf_3, \cdots, sf_k \}$ to  ML classifiers}
 }
 \ENDFOR
\FOR{each $sf_j$ $\in$ $SF$}
    {\STATE Apply class balancing
    \RETURN Class balanced data}
\ENDFOR
\FOR{each class balanced data $\in$ $SF$}
    {\STATE feed to ML classifiers}
\ENDFOR
\end{algorithmic}
\vspace*{.1cm}
{\bf Output:}
\smallskip $y \in classification \ output       \{c_1, c_2, \cdots, c_m\}$
\end{algorithm}


\begin{table*}[!htbp]
\centering
\caption{Classification performance metrics using the Random Forest algorithm with various frameworks}
\label{tab:Main-results}
\resizebox{1.7\columnwidth}{!}{%
\begin{tabular}{|c|c|c|c|c|c|c|}
\hline
\textbf{Framework} & \textbf{No. of Classes} & \textbf{Precision} & \textbf{Recall} & \textbf{$F_1$ score} & \textbf{Accuracy} & \textbf{Cohen Kappa Score} \\ \hline
\multirow{3}{*}{\textbf{FW1:Base}} & 34 & 0.7051 & 0.7866 & 0.7134 & 0.9918 & 0.991 \\ \cline{2-7} 
 & 8 & 0.7026 & 0.9022 & 0.7152 & 0.9944 & 0.9873 \\ \cline{2-7} 
 & 2 & 0.9671 & 0.9648 & 0.966 & 0.9968 & 0.932 \\ \hline
\multirow{3}{*}{\textbf{FW2:Base + ROS}} & 34 & 0.7031 & 0.8128 & 0.7119 & 0.9917 & 0.99 \\ \cline{2-7} 
 & 8 & 0.7018 & 0.9078 & 0.7138 & 0.9943 & 0.9871 \\ \cline{2-7} 
 & 2 & 0.9673 & 0.9645 & 0.9659 & 0.9968 & 0.9319 \\ \hline
\multirow{3}{*}{\textbf{FW2:Base + BRFC}} & 34 & 0.7014 & 0.786 & 0.7097 & 0.9915 & 0.99073 \\ \cline{2-7} 
 & 8 & 0.7049 & 0.9077 & 0.7186 & 0.9945 & 0.9874 \\ \cline{2-7} 
 & 2 & 0.9668 & 0.9645 & 0.9656 & 0.9968 & 0.9313 \\ \hline
\multirow{3}{*}{\textbf{FW3:CFS}} & 34 & 0.7381 & 0.8159 & 0.755 & 0.9923 & 0.9915 \\ \cline{2-7} 
 & 8 & 0.781 & 0.9016 & 0.8175 & 0.9943 & 0.9869 \\ \cline{2-7} 
 & 2 & 0.9599 & 0.9595 & 0.9597 & 0.9962 & 0.9194 \\ \hline
\multirow{3}{*}{\textbf{FW3:RF\_MRMR}} & 34 & 0.7301 & 0.8248 & 0.7488 & 0.9923 & 0.9916 \\ \cline{2-7} 
 & 8 & 0.7725 & 0.9284 & 0.8168 & 0.994 & 0.9863 \\ \cline{2-7} 
 & 2 & 0.9588 & 0.9458 & 0.9522 & 0.9955 & 0.9044 \\ \hline
\multirow{3}{*}{\textbf{FW4: CFS + ROS}} & 34 & 0.7396 & 0.8153 & 0.7564 & 0.9924 & 0.9917 \\ \cline{2-7} 
 & 8 & 0.7807 & 0.899 & 0.8167 & 0.9943 & 0.9869 \\ \cline{2-7} 
 & 2 & 0.9601 & 0.9597 & 0.9599 & 0.9963 & 0.9198 \\ \hline
\multirow{3}{*}{\textbf{FW4: CFS + BRFC}} & 34 & 0.7423 & 0.8241 & 0.7603 & 0.9923 & 0.9916 \\ \cline{2-7} 
 & 8 & 0.7825 & 0.9029 & 0.8186 & 0.9943 & 0.9869 \\ \cline{2-7} 
 & 2 & 0.9599 & 0.9592 & 0.9596 & 0.9962 & 0.9192 \\ \hline
\multirow{3}{*}{\textbf{FW4:RF\_MRMR + ROS}} & 34 & 0.7263 & 0.82 & 0.7447 & 0.9923 & 0.9915 \\ \cline{2-7} 
 & 8 & 0.7724 & 0.928 & 0.8168 & 0.994 & 0.9863 \\ \cline{2-7} 
 & 2 & 0.9595 & 0.9454 & 0.9523 & 0.9955 & 0.9047 \\ \hline
\multirow{3}{*}{\textbf{FW4:RF\_MRMR + BRFC}} & 34 & 0.7283 & 0.8313 & 0.74709 & 0.9923 & 0.9916 \\ \cline{2-7} 
 & 8 & 0.7723 & 0.9262 & 0.8155 & 0.99403 & 0.9863 \\ \cline{2-7} 
 & 2 & 0.959 & 0.9457 & 0.9522 & 0.9955 & 0.9045 \\ \hline
\end{tabular}%
}
\end{table*}

\begin{figure*}[!htbp]
\begin{center}
\fbox{\includegraphics[width=0.95\textwidth]{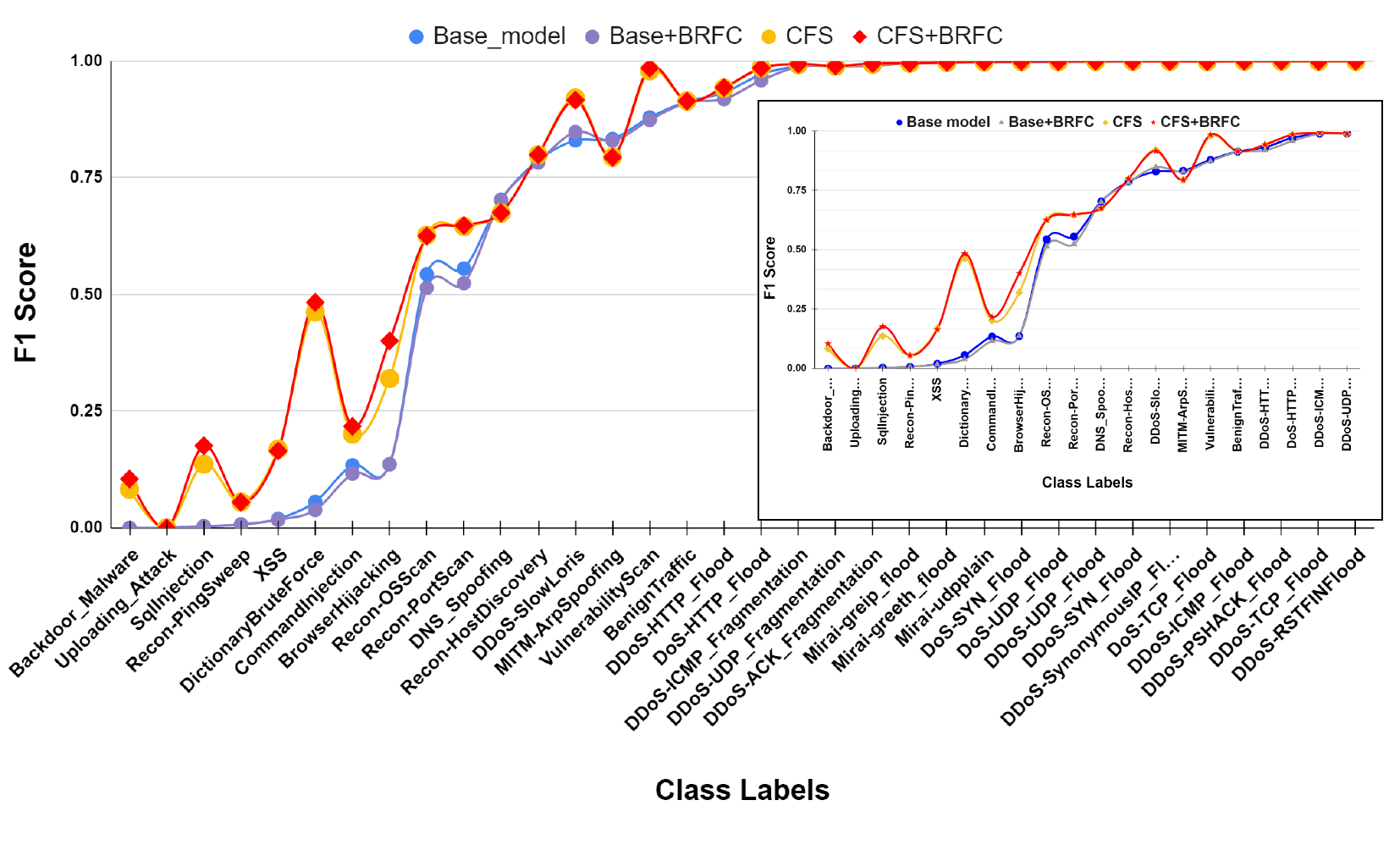}}
\end{center}
\caption{Proposed Performance Comparison in terms of $F_1$ Score } \label{fig:F1-score-compare-combined}
\end{figure*}

\section{Implementation} \label{ss:exp-details}
Following the feature selection and the class balancing stages, in the classification stage, the random forest algorithm is applied on the normalized set of the selected and balanced features from $CICIoT2023$ dataset. 
To demonstrate the efficacy of the proposed architecture, the following frameworks (FW) are implemented and compared. 
\begin{itemize}
    \item \textbf{FW1}: Base model \cite{ciciot2023}, consists of the original $46$ features from the CIC dataset analyzed with the random forest algorithm. 
    \item \textbf{FW2}: Consists of application of class balancing methods on the FW1. Specifically, the methods RoS and BRFC methods described in Sec.~\ref{subsec:ClassBalance} are applied in conjunction with FW1.  
    \item \textbf{FW3}: FW1 with feature selection methods in Sec. \ref{sec:MainMethod}.
    \item \textbf{FW4}: Proposed method with feature subset selection and class balancing methods as described in Sec.~\ref{sec:MainMethod} and Fig.~\ref{fig:architecture}.
\end{itemize}

\subsection{Assessment Scenarios}\label{ss:Assessment}
\begin{itemize}
    \item \textbf{Number of Categories:}  In each of the above mentioned experiments, the classification performance is 
    assessed for the following class label setups: 1) all the $34$ categories are considered 2) the events grouped to $8$ high level categories and 3) binary classification where normal event vs. attack is detected.  
    \item \textbf{Focus on Unsaturated Classes (USC):} 
    In this study, a number of events (classes) are identified as unsaturated classes (USC) for which the $F_1$ score $< 0.99$ with the base model. For the complementary set of classes, saturated classes (SC) ( $F_1$ score $\geq 0.99$) the performance may not improve as much and will be presented in the results section. Hence the focus is on the USC where the performance can be improved with various methods. The performance improvement for this set of classes is analyzed based on the class specific $F_1$ score.
\end{itemize}

\section{Results and Analysis} \label{ss:Result-analysis}
As discussed in the Sec. \ref{ss:Assessment}, the number of classes considered for categorization is dependent on the nature of attacks.
Based on the kind of attacks, $33$ attacks are grouped into the following seven categories: DoS, DDoS, Web-based, Recon, spoofing, Mirai, and bruteforce. 
Since each of these categories are attack related, for the binary classification, they are combined as a unifed attack class~\cite{ciciot2023}. The following analyses is based on the results for these three levels of categorization. 

\subsection{Analysis on ML Frameworks}
In the experiment 1, the results obtained from the four frameworks with variations as discussed in Sec. \ref{ss:exp-details} with the Random Forest algorithm are presented in Table \ref{tab:Main-results}. Note the Random Forest algorithm is chosen as it outperforms other models as discussed in the \cite{ciciot2023}.
From  Table \ref{tab:Main-results}, there is a minor improvement in the FW2 Base + RoS in comparison with FW1. There is 2.62\% improvement in recall score with ROS. However, there is a little improvement in the precision, $F_1$ score, accuracy and kappa. Thus the Balancing methods alone may not provide significant improvement.

\textbf{FW3 vs. FW1:}
In the FW3, the two feature subset selection techniques are applied separately on the dataset. In the first approach, the CFS identified $6$ best features out of $46$ features. The results of FW3: CFS, i.e., classification metrics corresponding to the RF classifier on these 6 CFS features are shown in the Table.~\ref{tab:Main-results}. In the second feature selection approach, the top 25 features from the RF with RFE, another set of $25$ best features using the MRMR technique are selected. Next an intersection of these subsets consisting of $11$ features is chosen. The results of RF model obtained from RF\_MRMR are listed in Table.~\ref{tab:Main-results}. For the FW3 CFS, there is $3.3$\%, $2.93$\% and $4.16$\% overall improvement on the base model FW1 in terms of precision, recall, $F_1$ score respectively. Note the recall for FW3 RF\_MRMR's is better than that of FW3 CFS. The accuracy and kappa measures are almost same for these two variatoins. From this comparison, it is observed that the FW3: CFS outperforms than the FW3: RF\_MRMR. 

\textbf{FW4 vs. Other Frameworks}
For the FW4 (proposed method), the results obtained from the four combinations of the feature sets and  the class balancing methdos are  listed in Table.~\ref{tab:Main-results}. In comparison with FW1, FW4: CFS+ROS outperforms in terms of precision and $F_1$ score. With FW4: CFS+ROS, there is an improvement of 3.45\%,  2.87\%, and 4.3\% in comparison with the base model in terms of precision, recall, and $F_1$ score respectively. FW4 RF\_MRMR+ROS recall is  0.47\% better than the CFS+ROS. The accuracy and kappa measures are similar for these variations. Again in comparison with FW1, Fw4: CFS+BRFC is outperforming in terms of precision and $F_1$ score. With FW4: CFS+BRFC, there is an improvement of 3.72\%,  3.75\%, and 4.69\% on the base model FW1 in terms of precision, recal, and $F_1$ score respectively. The recall for  FW4: RF\_MRMR+BRFC is  0.72\% better than that of the FW4: CFS+BRFC. From these observations, the FW4: CFS+BRFC approach performs well with only $6$ features compared to that of the base model and other frameworks.

\subsection{Analysis on Unsaturated Classes}
\begin{figure}[!htpb]
\begin{center} 
\fbox{\includegraphics[width=0.98\columnwidth]{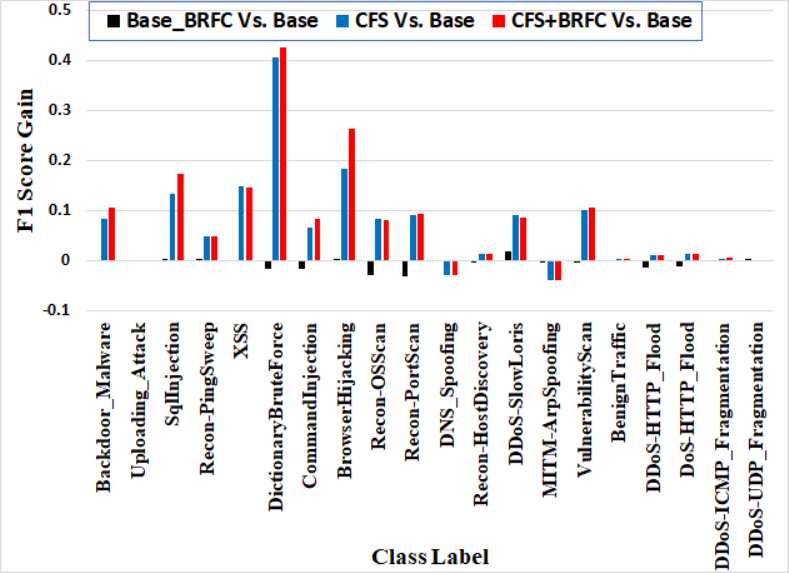}}
\end{center}
\scriptsize{\textbf{Observations:} The Average gain vs. $F_1$ for the frameworks are given as: i) FW2 -0.6268\%., ii)  FW3:CFS is 7.0416\%. and iii)  FW4:CFS+BRFC is 7.9297\%.}
\caption{Gain in $F_1$ Score of Unsaturated classes} \label{fig:F1-score-compare-gain}
\end{figure}

Fig. \ref{fig:F1-score-compare-combined} illustrates the $F_1$ scores of the chosen variations among the four frameworks (1). FW1; (2). FW2: BRFC; (3). FW3: CFS and 
(4). FW4: CFS + BRFC. The $F_1$ score is generally improving in the FW3 and FW4 in comparison with FW1. Note that the set of classes following \textit{DDoS\_ACK\_Fragment}, have $F_1$ score of atleast $0.99$, which, based on the definition of USC in the Introduction, has become saturated classes. The rest of the classes, on the left of \textit{DDoS\_ACK\_Fragment}, constitutes the USC. The focus in terms of $F_1$ gain is on this set of USC. The inset in the Fig. \ref{fig:F1-score-compare-gain} clearly shows the $F_1$ of the USC for the above mentioned four framework variations. For unsaturated classes, the average gain in $F_1$ score for a framework with respect to FW1 is given in Fig. \ref{fig:F1-score-compare-gain}. A maximum gain of $7.9$\% is achieved with FW4: CFS+BRFc. Hence the CFS plays a critical role in improving the $F_1$ of the unsaturated classes by $7.04$\%. Its combination with BRFC class balancing method provides a further improvement of 0.9\%.

\subsubsection{Relation to sample size} 
There is an intersection between the unsaturated classes and the minority classes (classes with relatively small sample size). For instance, the sample size of $DictionaryBruteForce$, a minority class is $13064$. For this class, the $F_1$ score for the FW1 is 5.6\%, the FW3: CFS it is $46.27$\%, FW4: CFS+BRFC is $48.31$\%. Thus FW4 improves by $42.71$\% with respect to FW1 as shown in Fig. \ref{fig:F1-score-compare-gain}. Similarly, other minority classes $BrowserHijacking$ and $SqlInjection$ have sample sizes $5859$ and $5245$ respectively. i.e., nearly 0.01\% of the overall sample size. The $F_1$ scores for FW1 are 13.62\% are 0.3\%  and FW4: CFS+BRFC are  $40.06$\% and 17.6\%. Hence, as shown in Fig. \ref{fig:F1-score-compare-gain}, the $F_1$ gains for these two classes are  $26.44$\% and 
$17.3$\%. 

\subsubsection{Importance in Intrusion Detection}
The results presented in Fig. \ref{fig:F1-score-compare-gain} illustrate the improvement in the $F_1$ score for 
unsaturated classes. The $F_1$ score in the security applications is an important metric. Following observations illustrate the importance of improved prediction of specific attack classes: a) For instance, identifying $DictionaryBruteForce$ attacks in IoT devices is of utmost importance for ensuring the security and integrity of the devices, protecting user data, and ensuring the overall safety of the IoT ecosystem.  b) Further, it is crucial to identify and address $BrowserHijacking$ in IoT devices to safeguard security, privacy, and data of users while preserving the integrity and reputation of device manufacturers and service providers. c)  Additionally, identifying $SqlInjection$ vulnerabilities in IoT devices also plays a crucial role in safeguard data, providing device integrity, protect networks, and preserve the trust of consumers and stakeholders in the IoT ecosystem. It is an essential step toward establishing a secure and resilient IoT infrastructure.
 
\section{Conclusion and Future Work}
In this paper, we studied the impact of feature selection and class balancing techniques on machine learning algorithms on the $CICIoT2023$ dataset. We analysed the performance of the proposed $IIDS$ with ML algorithms. We compared our results with the base model, and the proposed IIDS improves $3.72$\%, $3.75$\%, and $4.69$\% in Precision, Recall, and $F1$-score, respectively. In addition, with unsaturated classes analysis, we obtained a significant improvement of $7.9$\% compared to the base model. Finally, it is concluded that the combination of feature selection with CFS and the class balancing with BRFC techniques outperformed the other frameworks. \textbf{Future work:} In future work, we aim to extend our study  on the feature selection and class balancing techniques and how they influence the ML-based models for imbalanced datasets particularly in intrusion detection systems for IoT applications.

\bibliographystyle{ieeetr}
\bibliography{ieeecict2023}


\end{document}